\documentclass[aps,prl,twocolumn,hyperlinks,superscriptaddress,floatfix,showpacs,citeautoscript,longbibliography,hyperlinks]{revtex4-1}
\usepackage{amsmath}
\usepackage{amssymb}
\usepackage{amsfonts}
\usepackage{dsfont}
\usepackage{color}
\usepackage{nicefrac}
\usepackage[autostyle]{csquotes}
\usepackage[colorlinks=true,citecolor=blue]{hyperref}
\usepackage{graphicx,graphics}
\usepackage{siunitx}
\usepackage{braket}
\usepackage{float}
\usepackage[dvipsnames]{xcolor}
\usepackage{xcolor}
\usepackage{comment}

\usepackage{soul}

\begin{document}

\title{Local and non-local two-electron tunneling processes in a Cooper pair splitter}

\author{Antti Ranni}
\email{antti.ranni@ftf.lth.se}
\affiliation{NanoLund and Solid State Physics, Lund University, Box 118, 22100 Lund, Sweden}
\author{Elsa T. Mannila}
\affiliation{Pico group, QTF Centre of Excellence, Department of Applied Physics, Aalto University School of Science, P.O. Box 13500, 00076 Aalto, Finland}
\author{Axel Eriksson}
\affiliation{NanoLund and Solid State Physics, Lund University, Box 118, 22100 Lund, Sweden}
\author{Dmitry S. Golubev}
\author{Jukka P. Pekola}
\affiliation{Pico group, QTF Centre of Excellence, Department of Applied Physics, Aalto University School of Science, P.O. Box 13500, 00076 Aalto, Finland}
\author{Ville F. Maisi}
\email{ville.maisi@ftf.lth.se}
\affiliation{NanoLund and Solid State Physics, Lund University, Box 118, 22100 Lund, Sweden}

\date{\today}

\begin{abstract}
We measure the tunneling rates and coupling coefficients for local Andreev, non-local Andreev and elastic cotunneling processes. The non-local Andreev process, giving rise to Cooper pair splitting, exhibits the same coupling coefficient as the elastic co-tunneling whereas the local Andreev process is more than two orders of magnitude stronger than the corresponding non-local one. Theory estimates describe the findings and explain the large difference in the non-local and local coupling arising from competition between electron diffusion in the superconductor and tunnel junction transparency.
\end{abstract}

\maketitle

Superconductors give rise to electronic transport via two-electron processes as Cooper pairs cross for example tunnel junctions~\cite{Tinkham2004}. The two-electron transport enables several functionalities used extensively nowadays in quantum technology to build for example superconducting qubits~\cite{Makhlin1999,Nakamura1999,Wallraff2004,Andersson2019,Arute2019}, Majorana fermions~\cite{Mourik2012,Albrecht2016} and the Cooper pair splitters~\cite{Recher2001,Lesovik2001,Beckmann2004,Russo2005,Yeyati2007,Hofstetter2009,Zimansky2009,Herrmann2010} that are in focus in this study. The two electrons of a Cooper pair, however, yield typically several alternative transport processes that may happen. Understanding the coupling coefficients for the different processes is crucial as the coefficients depend on each other~\cite{Falci2001}. The coefficients contain information about the geometry and materials involved in the transport. A common approach in the experiments is to study the energy dependence of the dominant transport process~\cite{Eiles1993,Greibe2011,Maisi2011,Russo2005,Zimansky2009,Hofstetter2011,Tan2015,Tan2021}. The comparison between the coupling coefficients of different processes has evaded measurements since it is often experimentally difficult to distinguish them from each other. In this letter, we use charge readout with two detectors to identify each tunneling event in a Cooper pair splitter~\cite{Ranni2021} and compare the strength of three two-electron tunneling processes. The Cooper pair splitter is an ideal device for this purpose as both local Andreev~\cite{Andreev1964,Blonder1982}, non-local Andreev~\cite{Byers1995,Deutscher2000} and elastic cotunneling~\cite{Averin1990} transfer electrons across the two junctions located in the near vicinity of each other. By measuring the tunneling rates for the three two-electron processes at zero energy cost and extracting the corresponding coupling coefficients from the rates, we expand the knowledge of physics behind two-electron tunneling. Our experimental results, supported by theory predictions, demonstrate that the elastic cotunneling has the same coupling strength as the non-local Andreev tunneling and that the local Andreev process is two orders of magnitude stronger than the non-local one in our structure as a result of limited electron diffusion in the superconductor.

\begin{figure}
    \centering
	\includegraphics[width=\columnwidth]{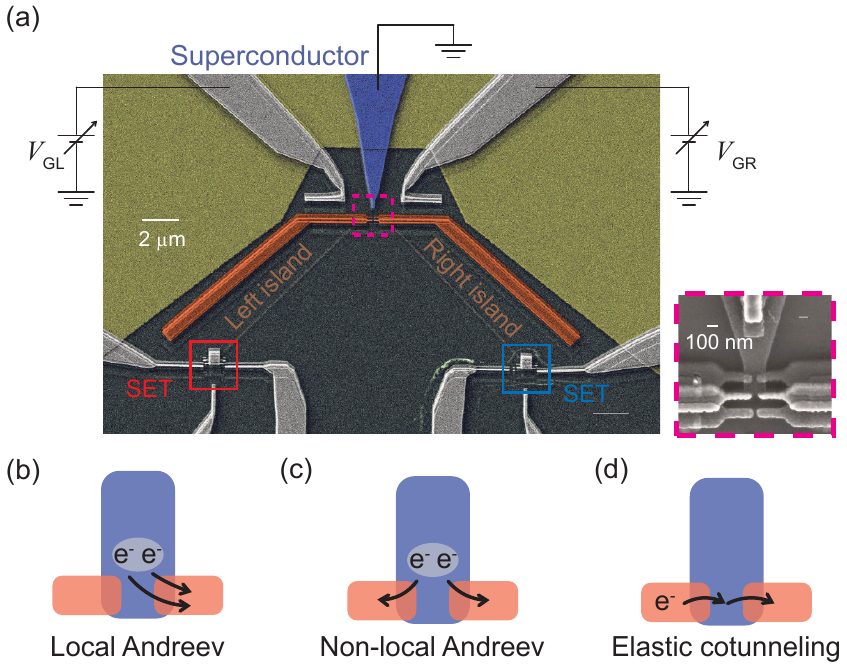}
	\caption{(a) Scanning electron micrograph of the studied device consisting of two copper islands (coloured orange) coupled to a superconducting aluminum electrode (in blue) via tunnel junctions~\cite{Ranni2021}. The inset on the right shows a zoom-in of the junctions. The electronic populations on the islands are controlled by the voltages $V_{\mathrm{GL}}$ and $V_{\mathrm{GR}}$ applied to the gate electrodes. 
	(b) Local Andreev tunneling process where a Cooper pair tunnels from the central superconductor into the right island.
	(c) A Cooper pair splitting where the two electrons forming the pair tunnel into separate metallic islands in non-local Andreev tunneling event.
	(d) Elastic cotunneling process where an electron moves from one island into the other via a virtual state in the superconductor.
	\label{fig1} 
		}
\end{figure}

We investigate electron tunneling in a recently realized Cooper pair splitter device~\cite{Ranni2021} where electron tunneling takes place between a superconductor and two normal metallic islands as depicted in Fig.~\ref{fig1}(a). Two single-electron transistors (SETs)~\cite{Schoelkopf1998,Gustavsson2009,Pekola2013} act as charge detectors observing the instantaneous charge state of both islands and thus resolving tunneling events as they occur. This charge counting technique yields access to the tunneling rates of the three two-electron processes allowing us to extract the couplings of these processes.

\begin{figure}[t]
    \centering
	\includegraphics[width=.98\columnwidth]{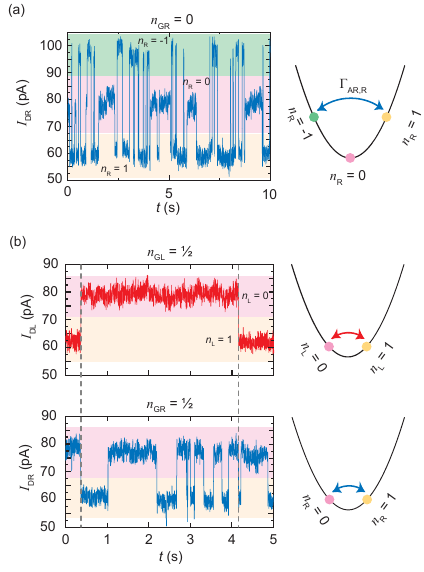}
	\caption{
	(a) The detector signal $I_{\mathrm{DR}}$ as a function of time $t$ revealing three charge states $n_{\mathrm{R}}$ on the right island corresponding to the energy diagram on the right yielding $\Gamma_\mathrm{AR,R}$.
	(b) The detector currents $I_{\mathrm{DL}}$ and $I_{\mathrm{DR}}$ as a function of time, recorded simultaneously. The dashed lines highlight non-local tunneling events. The islands are tuned to have the charge states $n_{\alpha} = 0,1$ degenerate in energy as presented in the energy diagrams. $n_{\mathrm{G\alpha}}$ is the normalized offset charge. These data was recorded at the base temperature of $T = 10$ mK in a dilution refrigerator.
	\label{fig2}
		}
\end{figure}

Local Andreev tunneling is schematically displayed in the diagram in Fig.~\ref{fig1}(b) where the two electrons forming a Cooper pair tunnel from the superconductor into either of the islands. To measure the local Andreev tunneling rate $\Gamma_\mathrm{AR, R}$ on the right island, we follow the proceduce of Ref.~\citealp{Maisi2011}: One of the charge states denoted with $n_R = 0$ excess electrons, is tuned with the gate voltage $V_{\mathrm{GR}}$ to be lowest in energy as presented in Fig.~\ref{fig2}(a). This configuration makes the charge states $n_R = \pm 1$ to be degenerate as shown in the energy diagram and local Andreev tunneling takes place between these states as seen in the measured time trace in Fig.~\ref{fig2}(a). By determining the number of tunneling events per time spent in the initial state  (Ref.~\citealp{Maisi2011}), we obtain the local Andreev tunneling rates $\Gamma_\mathrm{AR, R}^{\mathrm{in}} = 6.2$ Hz into and $\Gamma_\mathrm{AR, R}^{\mathrm{out}} = 6.5$ Hz out from the right island. Here we used the time window of $4$ ms corresponding to the detector risetime to determine if two consecutive events are from the same process or not. Since the rates are essentially the same, the tunneling indeed takes place without energy cost. Measuring with the other detector and tuning $V_{\mathrm{GL}}$ instead, we obtain similarly the local Andreev tunneling rates $\Gamma_\mathrm{AR, L}^{\mathrm{in}} = 61$ Hz and $\Gamma_\mathrm{AR, L}^{\mathrm{out}} = 55$ Hz on the left side. Interestingly, the left side has an order of magnitude larger rates despite the junctions have the same area (see inset of Fig.~\ref{fig1}(a)) and are made in the same process round very close to each other. The difference in the rates arise likely from differences in barrier thicknesses changing the channel transparencies~\cite{Maisi2011,Greibe2011,Aref2014}.

\begin{figure}[t]
	\centering
	\includegraphics[width=.98\columnwidth]{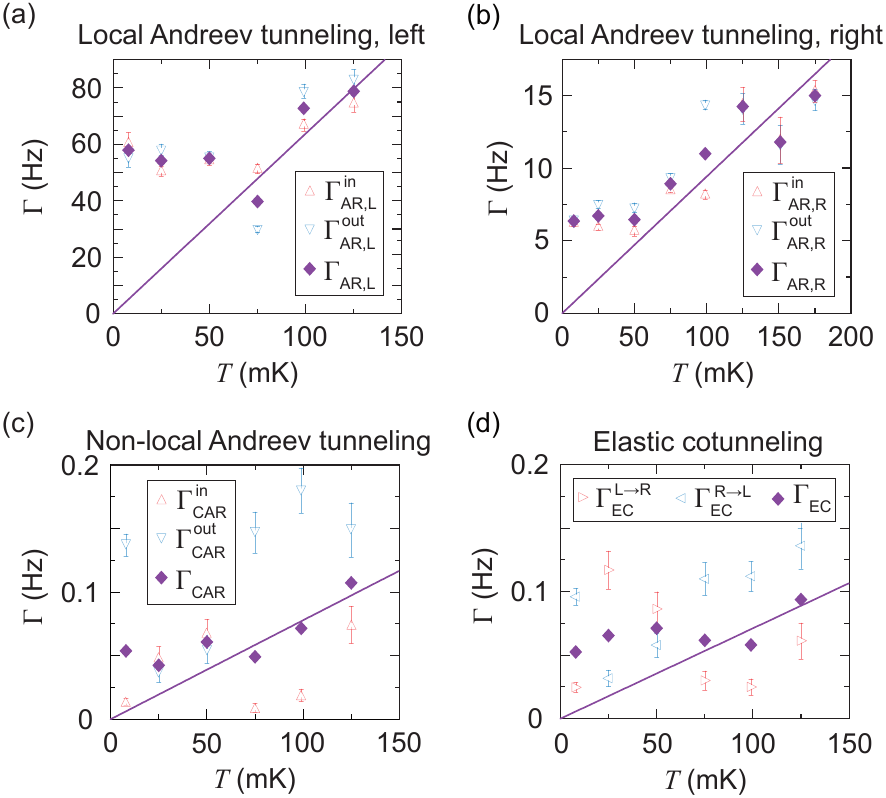}
	\caption{(a,b) Local Andreev tunneling rates at the left and the right junction, respectively. The open red triangles denote tunneling in, open blue triangles tunneling out of an island and the solid purple diamonds the logarithmic averages. The solid purple line is a linear fit to the logarithmic averages above the saturation, $T>\SI{50}{mK}$. 
	(c) Non-local Andreev tunneling rates in the similar manner. 
	(d) Elastic cotunneling rates. Here open red and blue triangles denote tunneling from left to right and vice versa.
	\label{fig3}
		} 
\end{figure}

The non-local two-electron processes are illustrated in Figs.~\ref{fig1}(c),(d). Panel~(c) presents non-local Andreev tunneling where the electrons forming a Cooper pair split into separate islands and panel~(d) elastic co-tunneling where an electron moves from one island to the other via the superconductor. To determine the tunneling rates $\Gamma_\mathrm{CAR}$ and $\Gamma_\mathrm{EC}$ for these processes at zero energy cost, the charge states $n=0$ and $n=1$ are tuned to degeneracy on both islands with the gate voltages $V_{\mathrm{GL}}$ and $V_{\mathrm{GR}}$. The measured time traces, shown in Fig.~\ref{fig2}(b), have equal occupancy of the charge states and hence are at equal energy as indicated by the energy diagrams. The non-local processes are then identified from the transitions as described in Ref.~\citealp{Ranni2021} as transitions within a time window of $1.5$ ms on both detectors. Here the time resolution is limited by the noise jitter between the two detectors instead of the detector rise time. At time $t = 0.3$~s, the left island loses an electron and the right one obtains one. Thus we had an elastic cotunneling from the left island to the right one. Similarly, at $t = 4.2$~s both islands obtain simultaneously an electron resulting in from Cooper pair splitting. We determine the tunneling rates
$\Gamma_{\mathrm{CAR}}^{\mathrm{in}}=\SI{14}{mHz}$ for Cooper pair splitting, $\Gamma_{\mathrm{CAR}}^{\mathrm{out}}=\SI{140}{mHz}$ for Cooper pair forming, $\Gamma_{\mathrm{EC}}^{\mathrm{L}\rightarrow\mathrm{R}}=\SI{25}{mHz}$  and $\Gamma_{\mathrm{EC}}^{\mathrm{R}\rightarrow\mathrm{L}}=\SI{96}{mHz}$ similarly as the local Andreev rates as the number of events divided by the time spent in the initial state. Despite of keeping the charge states $n = 0,1$ degenerate, the rates in the two directions, in vs. out, and $\mathrm{L}\rightarrow\mathrm{R}$ vs $\mathrm{R}\rightarrow\mathrm{L}$, are not equal. Such a difference arises from a finite energy gain $\delta E$ in one direction that appears as an energy cost in the opposing direction decreasing the rate. The rates for a two-electron process at $\delta E = 0$ may, however, be still determined with logarithmic averages of the rates in the two directions of the process e.g. for non-local Andreev as $\Gamma_{\mathrm{CAR}} = (\Gamma_{\mathrm{CAR}}^{\mathrm{in}}-\Gamma_{\mathrm{CAR}}^{\mathrm{out}})/(\mathrm{ln}(\Gamma_{\mathrm{CAR}}^{\mathrm{in}})-\mathrm{ln}(\Gamma_{\mathrm{CAR}}^{\mathrm{out}}))$. Figure~\ref{fig3} summarizes the two-electron tunneling rate measurements where we repeated the experiment at varying bath temperature $T$. 

The second-order perturbation theory~\cite{Hekking1994} yields a general expression for the above three two-electron rates as
\begin{equation}
    \Gamma_{\mathrm{2e}}(\delta E, T)=\gamma\frac{\delta E/k_{\mathrm{B}}T}{1-\mathrm{e}^{-\delta E/k_{\mathrm{B}}T}}k_{\mathrm{B}}T,
    \label{eq:2eRates}
\end{equation}
where $\gamma$ is a coupling constant and $k_{\mathrm{B}}$ the Boltzmann constant. All two-electron processes follow the same functional dependence of the energy cost $\delta E$ per thermal energy $k_{\mathrm{B}}T$. The energy cost $\delta E$ itself, and the coupling $\gamma$, are however not the same for different processes. The charging energy difference between the initial and final state sets $\delta E$ for the two-electron processes as $\delta E_\mathrm{AR,\alpha}^{\mathrm{in}}=4E_{\mathrm{C}}\; n_{\mathrm{G\alpha}}$ (with $\alpha=\mathrm{L,R}$), $\delta E_\mathrm{CAR}^{\mathrm{in}}=2E_{\mathrm{C}}(n_{\mathrm{GL}}+n_{\mathrm{GR}}-1)$ and $\delta E_\mathrm{EC}^{\mathrm{L\rightarrow R}}=2E_{\mathrm{C}}(n_{\mathrm{GR}}-n_{\mathrm{GL}})$. The costs to opposite tunneling directions are the same but with opposite signs. Here $E_{\mathrm{C}}$ is the charging energy of individual identical islands and $n_{\mathrm{G}\alpha}$ the normalized offset charge controlled by $V_{\mathrm{G}\alpha}$~\cite{Lafarge1991}. The cost vanishes when the initial and final state of the process are at the same energy as depicted in Fig.~\ref{fig2}. Equation~(\ref{eq:2eRates}) acquires in this case a simple form $\Gamma_{\mathrm{2e}}(0,T)=\gamma\; k_{\mathrm{B}}T$. 


The coupling terms for the three two-electron processes read as  
\begin{equation}
    \begin{split}
    \gamma_{\mathrm{AR}, \alpha} &= \frac{1}{8e^{2}R_{\mathrm{T\alpha}}^{2}}\frac{R_{\mathrm{K}}}{\mathcal{N}_{\alpha}}, \\
    \gamma_{\mathrm{CAR}} &= \gamma_{\mathrm{EC}} = \frac{\mathrm{e}^{-l/\xi}}{2e^{2}R_{\mathrm{TL}}R_{\mathrm{TR}}}R_{\mathrm{S}}.
    \end{split}
    \label{eq:gamma}
\end{equation}
Here $\mathcal{N}_{\alpha}=A_{\alpha}/A_{\mathrm{ch},\alpha}$ is the effective number of the conduction channels in a junction $\alpha=\mathrm{L,R}$ with a junction area $A_{\alpha}$ and effective conduction channel area $A_{\mathrm{ch}, \alpha}$~\cite{Maisi2011}. $R_{\mathrm{T}\alpha}$ is the junction resistance, $R_{\mathrm{K}}\equiv h/e^{2}$ the so-called resistance quantum, $e$ the elementary charge, $l$ the distance between the two tunnel junctions, $\xi$ the superconducting coherence length and $R_{\mathrm{S}}$ the normal-state sheet resistance of the superconducting electrode measured over the superconducting coherence length. Interestingly, the coupling terms for non-local Andreev $\gamma_{\mathrm{CAR}}$ and elastic cotunneling $\gamma_{\mathrm{EC}}$ are identical according to theory. Non-local Andreev either splits or assembles Cooper pairs whereas elastic cotunneling does not involve pairing and takes place even in the absence of superconductivity. 

On the other hand, the local tunnel coupling $\gamma_\mathrm{AR}$ differs from $\gamma_\mathrm{CAR}$ and $\gamma_\mathrm{EC}$~\cite{Falci2001,Golubev2007,Brauer2010}. As seen from Eq.~(\ref{eq:gamma}), $\gamma_\mathrm{AR}$ depends on the number of conduction channels $\mathcal{N}_{\alpha}$ in the junction and the junction resistance $R_{\mathrm{T\alpha}}$ in relation to the resistance quantum $R_{\mathrm{K}}$~\cite{Averin2008,Maisi2011,Aref2014}. The non-local processes depend instead on the total junction resistances versus quasiparticle diffusion away from the junction area set by $R_{\mathrm{S}}$~\cite{Golubev2019} but not on $\mathcal{N}_{\alpha}$ nor $R_{\mathrm{K}}$. In addition, the non-local processes have an exponential suppression $\mathrm{e}^{-l/\xi}$ for increasing distance $l$ between the two junctions. Local Andreev is free of this suppression as the process takes place across a single junction.

\begin{figure}[t]
	\centering
	\includegraphics[width=.98\columnwidth]{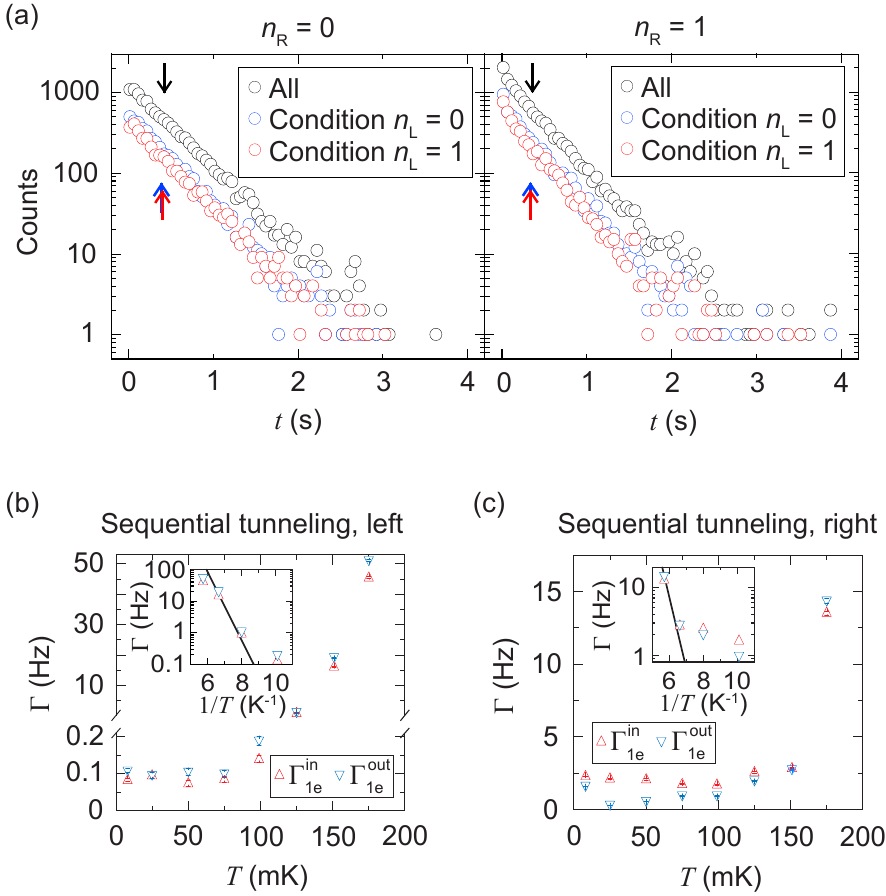}
	\caption{(a) Lifetime $t$ distribution of charge states $n_{\mathrm{R}}=0$ and 1 on the left and right graph, respectively. The black circles are lifetimes without any condition set for the charge state on the left island. The blue and red circles show the lifetime distribution on right island with the condition that the left island is in $n_{\mathrm{L}}=0$ or 1 state respectively during the whole lifetime $t$. The arrows denote the mean values of the distributions. (b,c) Sequential tunneling rates at the left and the right junction, respectively at zero energy cost. The open red (blue) triangles denote an electron tunneling into (out of) an island. The insets display the rates at four highest temperatures on a logarithmic scale against inverse temperature. The solid black lines are the theoretical models with tunnel resistances as free fitting parameters. \label{fig4}
		} 
\end{figure}

We turn now back to the experimental data and determine the couplings $\gamma$. Two-electron rates in Fig.~\ref{fig3} follow an increasing linear trend for $T>\SI{50}{mK}$. At lower temperatures the tunnel rates saturate to a fixed value arising from the saturation of electronic temperature in our dilution refrigerator at $\SI{50}{mK}$. The coupling terms are obtained by a linear fit (purple line) to rates at zero energy cost above the saturation in Fig.~\ref{fig3}. The extracted values are $\gamma_{\mathrm{AR,L}}=\SI{7.5}{/\micro eVs}$, $\gamma_{\mathrm{AR,R}}=\SI{1.1}{/\micro eVs}$, $\gamma_{\mathrm{CAR}}=\SI{9.0e-3}{/\micro eVs}$ and $\gamma_{\mathrm{EC}}=\SI{8.3e-3}{/\micro eVs}$. 

Based on the above fits, $\gamma_{\mathrm{CAR}}$ and $\gamma_{\mathrm{EC}}$ are equal within the experimental accuracy, as the theory of Eq.~(\ref{eq:gamma}) predicts. The result also implies that there is no significant capacitive coupling between the islands since with such coupling, the elastic cotunneling rates would become higher than the non-local Andreev rates as splitting a Cooper pair to separate islands would require additional energy to charge the two islands by one electron each~\cite{Yeyati2007,Beckmann2007}. This conclusion is supported by the lifetime distributions for the charge states $n_{\mathrm{R}}=0,1$ shown in Fig.~\ref{fig4} (a): The lifetime on the right island is independent of the occupancy on the left island. A capacitive coupling between the islands would favor energetically to have dissimilar electron numbers on the islands and hence decrease the lifetime for same electron numbers and increase it for differing electron numbers.  

The above fits also show that the coupling terms for local Andreev tunneling are two to three orders of magnitude larger than for non-local processes. From the expressions in Eq.~(\ref{eq:gamma}) we see that the possible explanations are the exponential suppression $\mathrm{e}^{-l/\xi}$ and how the terms $R_{\mathrm{K}}/\mathcal{N}$ and $R_{\mathrm{S}}$ compare to each other. To assess where the difference arises from, we estimate the parameters constituting the coupling terms of Eq.~(\ref{eq:gamma}) in the following manner: The sequential single-electron tunneling rates, measured also with the protocol described in Ref.~\citealp{Ranni2021}, are shown in Figs.~\ref{fig4}(b) and (c) across the left and the right junction. These yield us $R_{\mathrm{T\alpha}}$ since the sequential rates follow an exponential temperature dependence of $\Gamma_{\mathrm{1e}}(T)=\frac{1}{e^{2}R_{T\alpha}}\sqrt{2\pi \Delta k_{\mathrm{B}}T}\mathrm{e}^{-\Delta/k_{\mathrm{B}}T}$ without charging energy cost between $n_\alpha = 0,1$~\cite{Pekola2013}. Similarly to two-electron tunneling, the sequential rates exhibit saturation at low temperatures. However, the sequential rates saturate around $\SI{100}{mK}$ which is higher than for two-electron tunneling due to suppression of sequential tunneling by the superconductor energy gap $\Delta$. Sequential tunneling probes electron distributions in the superconductor and metallic islands above the gap energies whereas two-electron tunneling is sensitive to normal-state electrons around the Fermi energy. Another factor that might play a role in higher saturation temperature of sequential tunneling is that superconductors do not thermalize as efficiently as normal-state metals~\cite{Giazotto2006}. At the highest temperatures the measured sequential rates grow exponentially as shown in the insets of Figs.~\ref{fig4}(b,c). The superconducting gap for a 20 nm thin film is $\Delta = 210 \pm \SI{10}{\micro eV}$~\cite{Maisi2011,Aref2014,Mannila2019}. Hence, fitting $\Gamma_{\mathrm{1e}}(T)$ to the high temperature regime, as shown as the black lines in the inset, yields us the tunnel resistances 
$R_{\mathrm{TL}}=\SI{5}{M\Omega}$ and $R_{\mathrm{TR}}=\SI{50}{M\Omega}$ as the only free fitting parameters. Note that due to the strong exponential dependence, the $\pm \SI{10}{\micro eV}$ uncertainty in $\Delta$ yields a factor of two uncertainty to the resistance values.
 
Next, with the experimentally determined $\gamma_\mathrm{AR,\alpha}$ and Eq.~(\ref{eq:gamma}), we obtain the number of conduction channels as $\mathcal{N}_{\mathrm{L}} = 130$ and $\mathcal{N}_{\mathrm{R}} = 7$. With the junction areas $A_{\mathrm{L}}=\SI{85}{nm}\times \SI{80}{nm}$ and $A_{\mathrm{R}}=\SI{80}{nm}\times \SI{70}{nm}$ estimated from the scanning electron micrograph of Fig.~\ref{fig1} (a), the corresponding effective channel sizes are $A_{\mathrm{ch,L}}=\SI{50}{nm^2}$ and $A_{\mathrm{ch,R}}=\SI{790}{nm^2}$. The values for $A_{\mathrm{ch,\alpha}}$ are comparable to the earlier work of Refs.~\cite{Maisi2011,Aref2014} where $A_{\mathrm{ch}}=\SI{30}{nm^2}$ was reported for junctions fabricated with the same process.

Now we turn into the non-local Andreev process. We estimate $R_S$ from the normal-state resistance of a $\xi=\SI{100}{nm}$~\cite{Koski2011} long segment of the superconductor electrode next to the junctions. The width of the electrode near the junctions is $W=\SI{210}{nm}$ and the thickness $d=\SI{20}{nm}$ measured by a crystal monitor during metal evaporation. With the previously measured normal-state resistance of e-beam evaporated aluminium films in Ref.~\citealp{Knowles2012}, $\rho_{\mathrm{N}}=\SI{31}{n\Omega m}$, we obtain $R_{\mathrm{S}}=\rho_{\mathrm{N}}\frac{\xi}{Wd} = \SI{0.7}{\Omega}$. To determine the exponential suppression factor of $\mathrm{e}^{-l/\xi}$, we obtain the distance $l$ between the junctions from the scanning electron micrograph. The shortest distance between the junctions is $\SI{50}{nm}$ and the distance between the far edges of the junctions is $W=\SI{210}{nm}$. We apply the halfway $l=\SI{130}{nm}$. With the chosen values the exponential suppression becomes $\mathrm{e}^{-l/\xi}\approx 0.3$. If we instead applied the minimum or maximum distance in our approximation the value of the suppression would increase or decrease by a factor of two. With these independent estimates, we obtain from Eq. (\ref{eq:gamma}) the coupling estimate $\gamma_\mathrm{CAR} = \SI{3e-3}{/\micro eVs}$ which is in reasonable agreement with the experimentally determined values, considering the uncertainties in the parameter estimation. This parameter estimation allows us to conclude that the two to three orders of magnitude difference between the local and non-local Andreev process arises predominantly from the difference in the terms $R_{\mathrm{K}}/\mathcal{N}_{\alpha}$ and $R_{\mathrm{S}}$ respectively. In other words, the diffusion away via the superconductor suppresses the non-local process in the structure.

In conclusion, we used charge counting to determine the coupling coefficients for three two-electron tunneling processes relevant in Cooper pair splitters. Our experimental findings validate the theoretical prediction of $\gamma_{\mathrm{CAR}} = \gamma_{\mathrm{EC}}$. We also determined the coupling terms for local and non-local Andreev tunneling and found out that the non-local one is suppressed by more than two orders of magnitude because of competing diffusion in the superconductor.

We acknowledge Fredrik Brange, Christian Flindt, Martin Leijnse and Claes Thelander for fruitful discussions and the QuantERA project "2D hybrid materials as a platform for topological quantum computing", Swedish National Science Foundation (Dnr 2018-00099 and 2019-04111), NanoLund, Academy of Finland grant 312057 and QTF Centre of Excellence for financial support. We acknowledge the provision of facilities by Aalto University at OtaNano - Micronova Nanofabrication Centre.

\bibliography{main}

\end{document}